\documentclass[a4paper,11pt]{article}
\usepackage{jinstpub} % for details on the use of the package, please see the JINST-author-manual

\usepackage{subcaption}

\usepackage{cleveref}

\usepackage{lineno}
\title{\boldmath Evidence of Charge Multiplication in Thin $25 \mathrm{\mu m} \times 25 \mathrm{\mu m}$ Pitch 3D Silicon Sensors}

\author[a,1]{A. Gentry, \note{Corresponding author.}}
\author[b,c]{M. Boscardin,}
\author[a]{M. Hoeferkamp,}
\author[d]{M. Povoli,}
\author[a]{S. Seidel,}
\author[a]{J. Si,}
\author[c,e]{G.-F.~Dalla~Betta}
\affiliation[a]{Department of Physics and Astronomy, University of New Mexico,\\ 210 Yale Blvd.~NE, Albuquerque, NM 87106, U.S.A.}
\affiliation[b]{Fondazione Bruno Kessler, via Sommarive 18, Povo, Italy}
\affiliation[c]{TIFPA INFN, via Sommarive 14, Povo, Italy}
\affiliation[d]{SINTEF Digital, Smart Sensors and Microsystems, Gaustadalleen, 23C, 0373, Oslo, Norway}
\affiliation[e]{Department of Industrial Engineering, Universit\`a degli Studi di Trento and INFN,\\ via Sommarive 9, Povo, Italy}

\emailAdd{agentry2@unm.edu}

\abstract{Characterization measurements of $25~\mathrm{\mu m} \times 25~\mathrm{\mu m}$ pitch 3D silicon sensors are performed, for devices with active thickness of $150~\mu$m.  Evidence of charge multiplication caused by impact ionization below the breakdown voltage is observed in sensors operated at $-45~^\circ\mathrm{C}$. Small-pitch 3D silicon sensors have potential as high precision 4D tracking detectors that are also able to withstand radiation fluences beyond $10^{16}$~n$_{\rm eq}/$cm$^2$. This is applicable for use at future facilities such as the High-Luminosity Large Hadron Collider and the Future Circular Collider. Characteristics of these devices are compared to those of similar sensors of pitch $50~\mathrm{\mu m}\times 50~\mathrm{\mu m}$, showing comparable charge collection at low voltage, and acceptable leakage current, depletion voltage, breakdown voltage, and capacitance despite the extremely small cell size. The unirradiated $25~\mathrm{\mu m} \times 25~\mathrm{\mu m}$ sensors exhibit charge multiplication above about 90 V reverse bias, while, as predicted, no multiplication is observed in the $50~\mathrm{\mu m} \times 50~\mathrm{\mu m}$ sensors below their breakdown voltage. The maximum gain observed below breakdown is 1.33.}


\arxivnumber{1234.56789} % Only if you have one

\begin{document}
\maketitle
\flushbottom

\section{Introduction}

\label{sec:intro}
~
Silicon sensors with 3D electrodes have proven effective for radiation tolerance in high energy physics experiments \cite{3DBook}, and they will constitute a key component of the upgraded tracking detectors in both ATLAS and CMS at the High-Luminosity LHC (HL-LHC)~\cite{ATLAS:2010ojh,ATLAS:2017svb,CMS:2012sda}. The ATLAS Insertable B-Layer~\cite{daVia} incorporates 3D sensors of pitch $50~\mu{\rm m} \times 250~\mu{\rm m}$ and thickness $230~\mu{\rm m}$.  Smaller pixel sizes will provide better radiation hardness due to their low signal collection time relative to the charge trapping time. Sensors of pitch $50~\mathrm{\mu m}~\times~50~\mathrm{\mu m}$  and $25~\mathrm{\mu m}~\times~100~\mathrm{\mu m}$ are planned for operation at the HL-LHC; they have demonstrated good charge collection efficiency up to $\mathrm{3\times 10^{16}~n_{eq}/cm^2}$~\cite{Lange:2018paz}. 

3D sensors are also intrinsically fast detectors, owing to the small drift distance required for signal collection. They can be compared favorably with Low Gain Avalanche Detectors (LGADs), a promising technology for high resolution timing in moderate radiation conditions; installation of LGADs is planned for both ATLAS and CMS at the beginning of the HL-LHC era.  3D sensors with $50~\mathrm{\mu m} \times 50~\mathrm{\mu m}$ pitch have timing resolution $\sigma_t$ better than 50 ps, which is competitive with that of LGADs~\cite{Kramberger:2019ygz}.

ATLAS and CMS anticipate replacement of inner tracking layers during the lifetime of the HL-LHC due to challenges in withstanding the radiation dose anticipated during the full HL-LHC lifetime \cite{Gonella:2023uag}.  Studies have been conducted on the potential benefits of 4D tracking in these innermost layers, which could use small-pitch 3D sensors~\cite{Kramberger:2023qnh,ATLAS:2023tbj}.

Future experiments, for example those at the Future Circular Collider (FCC), will require even better timing precision due to increased pileup.  Timing measurements can be exploited at the FCC-ee to support particle identification of long-lived particles, and to facilitate pattern recognition~\cite{Bacchetta}.  The
FCC-hh would produce fluences greater than those at the HL-LHC by more than an order of magnitude~\cite{FCC:2018evy,FCC:2018vvp,Besana:2016srq}. Thus, 4D tracking devices with < 10 ps timing able to withstand > $10^{17} {\rm~n_{eq}/cm^2}$ are desirable.

A 3D pixel pitch below the $50~\mathrm{\mu m} \times 50~\mathrm{\mu m}$ or $25~\mathrm{\mu m} \times 100~\mathrm{\mu m}$ ones could improve functionality in both radiation hardness and timing resolution. Simulations of signal response in a $25~\mathrm{\mu m} \times 25~\mathrm{\mu m}$ 3D sensor predict timing resolution in the realm of $\sigma_t \sim 13$ ps~\cite{Kramberger:2023qnh}. Additionally, simulations have suggested the possibility of controlled charge multiplication below the breakdown voltage for this geometry, which could restore charge collection efficiency at extreme fluences \cite{Povoli:2019,DallaBetta:2022drc}.

We present here characterizations of a set of unirradiated $25~\mathrm{\mu m} \times 25~\mathrm{\mu m}$ pitch 3D sensors.  Measurements of their charge collection efficiency are included.  Comparisons to similar devices of larger pitch are provided. Evidence for charge multiplication in the unirradiated small pitch sensors is shown.

\section{Description of Devices}
\label{sec:Desc}

~
Two sets of sensors, of pitches  $25~\mathrm{\mu m} \times 25~\mathrm{\mu m}$ and $50~\mathrm{\mu m} \times 50~\mu$m, both with aluminum connections between readout columns placed at the center of 3D cells, were measured for comparison. These sensors were designed at the University of Trento and fabricated at Fondazione Bruno Kessler (FBK) using step-and-repeat (stepper) lithography, which allows the production to achieve 150 $\mu$m active thickness and pixel pitch down to $25~\mathrm{\mu m} \times 25~\mathrm{\mu m}$ \cite{Boscardin:2020two}. 

\begin{figure}[htbp]
\centering
    \begin{subcaptionblock}{0.4\textwidth}
        \includegraphics[width=\textwidth]{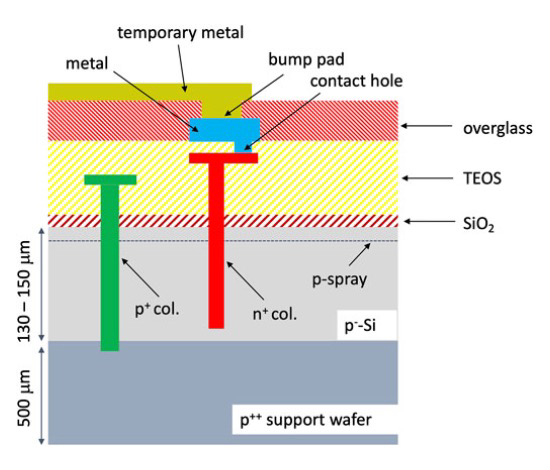}
        \caption{\label{fig:SensorXSection}}%
    \end{subcaptionblock}
%\qquad
    \begin{subcaptionblock}{0.25\textwidth}
        \includegraphics[width=\textwidth]{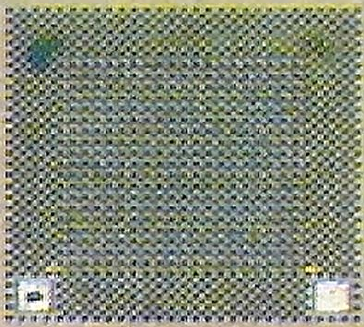}
        \caption{\label{fig:SensorPics}}%
    \end{subcaptionblock}
    \begin{subcaptionblock}{0.3\textwidth}
        \includegraphics[width=\textwidth]{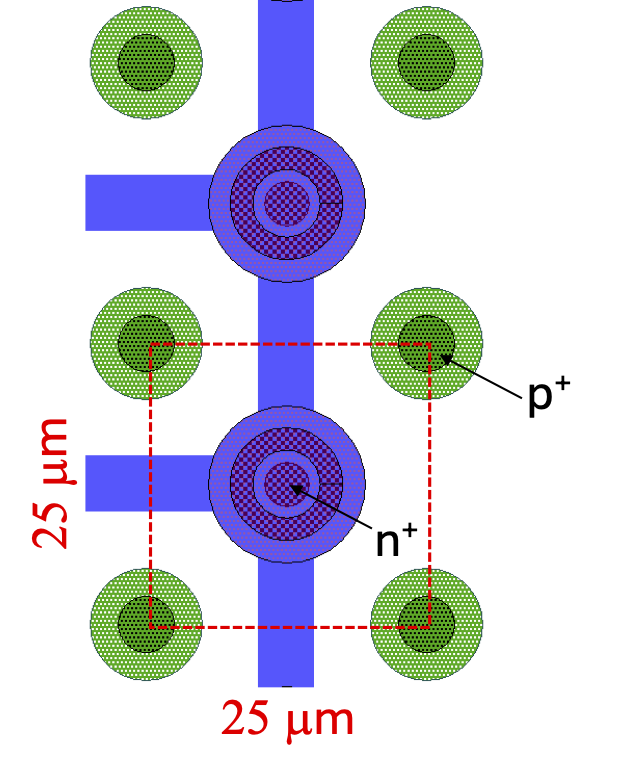}
        \caption{\label{fig:SensorDiagram}}%
    \end{subcaptionblock}
\caption{(a) A diagram of a cross section of the sensors under study. (b) An image of a $25~\mathrm{\mu m} \times 25~\mathrm{\mu m}$ pixel array with aluminum connections, for bench testing. (c) A diagram of the $25~\mathrm{\mu m} \times 25~\mathrm{\mu m}$ cell layout with aluminum connections.}.\label{fig:SensorImages}
\end{figure}

A diagram of the sensor design can be seen in Figure~\ref{fig:SensorXSection}. A p-type $150~\mu$m substrate with $\langle 100\rangle$ orientation and nominal resistivity in the range $5-10~\mathrm{k\Omega}$ is bonded to a $500~\mu$m low resistivity p-type support wafer and processed from the front side. A p-spray layer is used to assure surface isolation, and after this is applied, p-type columns are etched, penetrating the support wafer so that the sensor can be biased from the backside. Next, the n-type columns are etched, with an approximately $35~\mu$m gap between the column and the support wafer to avoid shorting and early breakdown. Both types of columns have a nominal diameter of $5~\mu$m, and the n-type columns have a diffusion profile on the order of $1~\mu$m. There are some non-uniformities in the column diameter from top to bottom, with the tip having up to about $1~\mu$m smaller diameter. Tetra-ethyl-ortho-silicate (TEOS) is used to protect the columns.  Contact holes are etched in the TEOS layer in order to make contact with the n-type columns.

Both the $25~\mathrm{\mu m} \times 25~\mathrm{\mu m}$ and the $50~\mathrm{\mu m} \times 50~\mathrm{\mu m}$ devices are $20\times 20$ arrays of pixels whose electrodes are connected by aluminum. An image of a sensor array and a detail of the $25~\mathrm{\mu m} \times 25~\mathrm{\mu m}$ layout are in Figures \ref{fig:SensorPics} and \ref{fig:SensorDiagram}, respectively.

\section{Measurement Setup}
\label{sec:setup}

~
Three types of measurements are presented: leakage current versus bias voltage (IV), bulk capacitance versus bias voltage (CV), and charge collection efficiency (CCE) in response to a $^{90}$Sr beta source.

The IV measurement measures the sensor leakage current and breakdown voltage. The sensor is placed inside a dark box on a Peltier-cooled thermal chuck and held at $20 \pm 0.5 ~^\circ$C. It is biased from the back side using a Keithley 237 source measure unit, and the leakage current is measured through a probe on the sensor pad that is connected to the column array.

The CV measurement shows the sensor depletion characteristic, which for a 3D sensor reflects the radial geometry of the electric field lines when the sensor is biased. The setup is similar to that of the IV measurements, with an HP4284A LCR meter and a bias isolation box used to measure the capacitance.

\begin{figure}[htbp]
\centering
    \begin{subcaptionblock}{0.50\textwidth}
        \includegraphics[width=\textwidth]{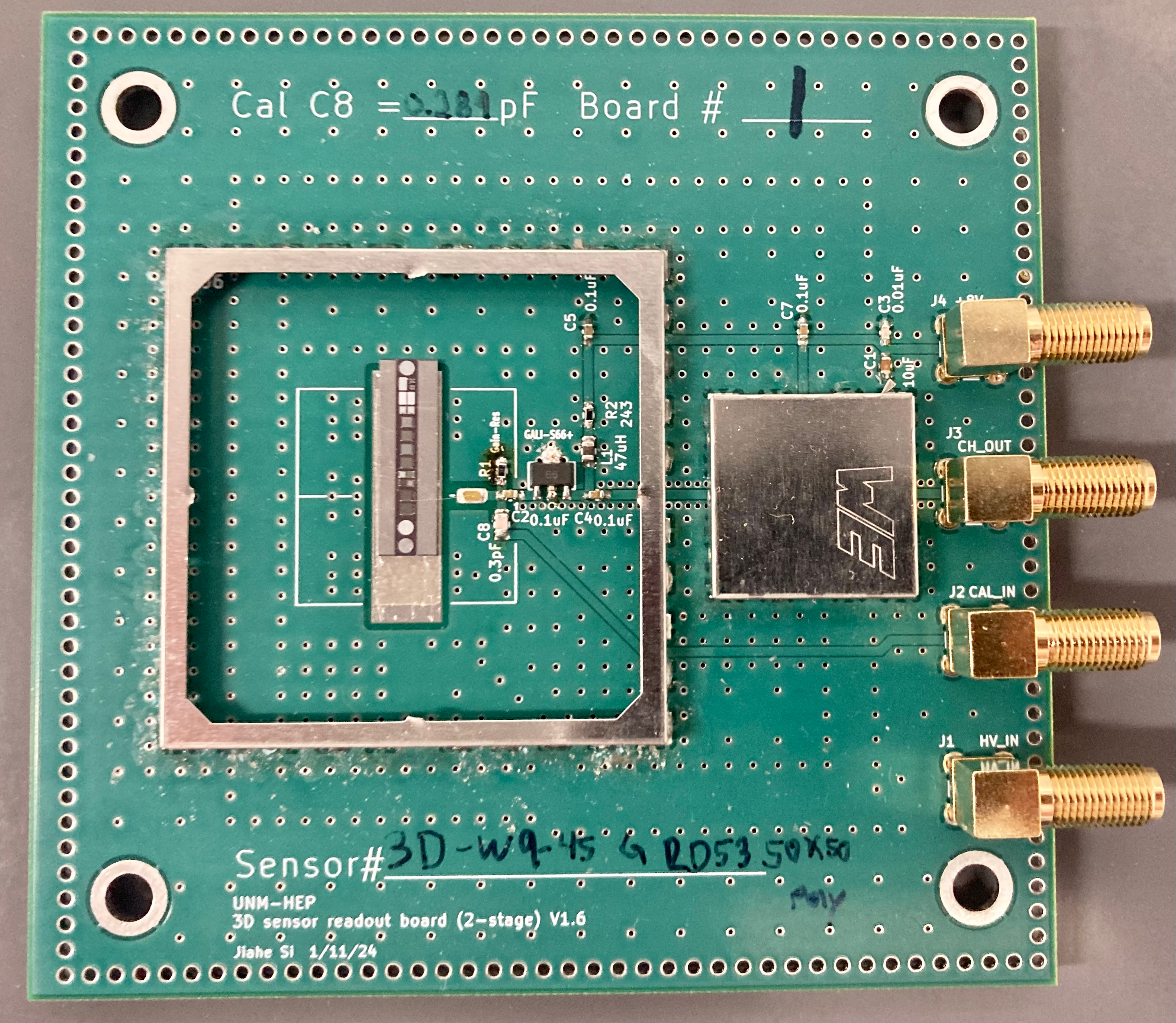}
        \caption{\label{fig:ReadoutBoard1}}%
    \end{subcaptionblock}
%\qquad
    \begin{subcaptionblock}{0.48\textwidth}
        \includegraphics[angle=270,width=\textwidth,trim={0 2cm 0 2cm},clip]{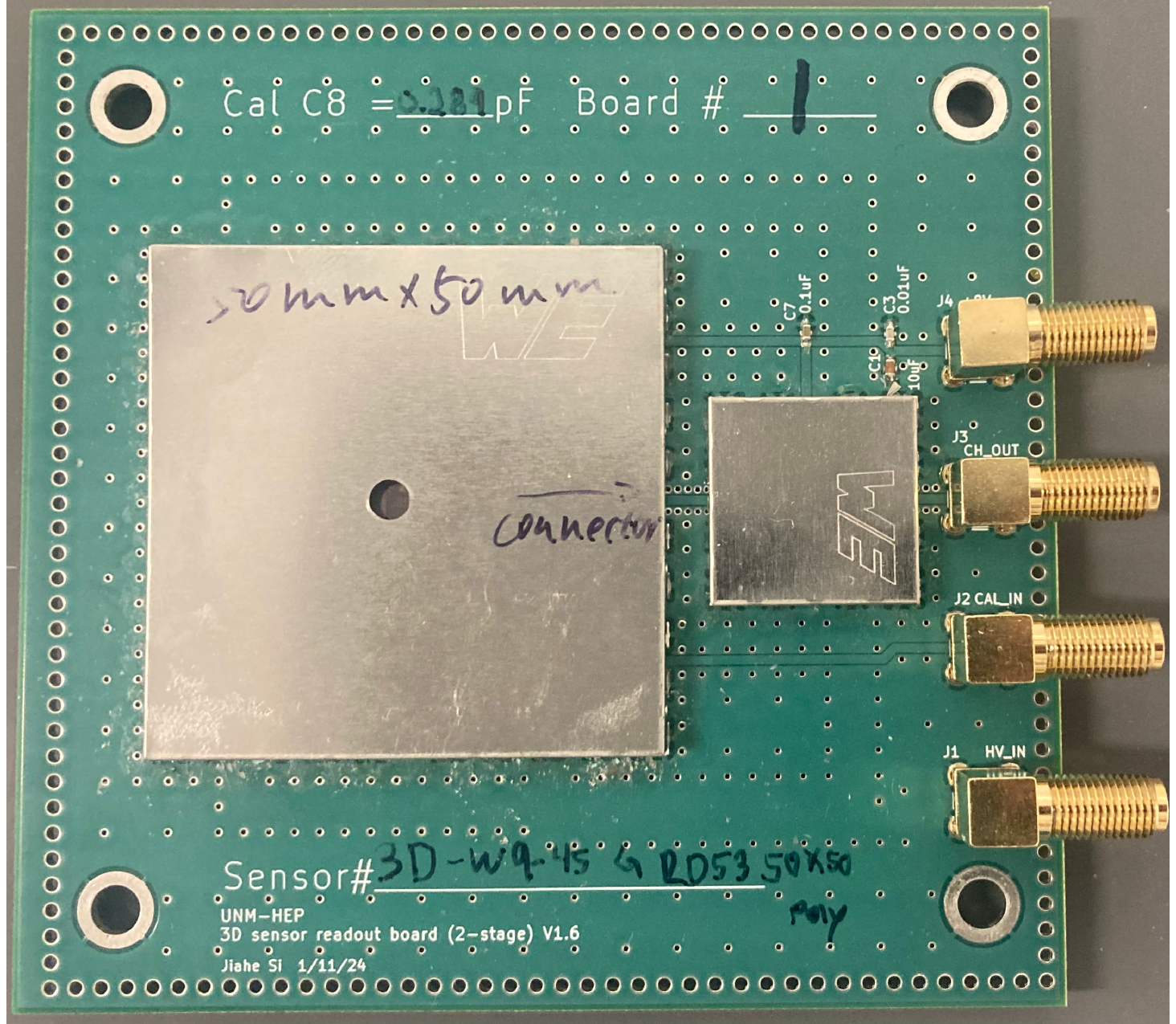}
        \caption{\label{fig:ReadoutBoard2}}%
    \end{subcaptionblock}
\caption{The readout board used in charge collection measurements, (a) without and (b) with the EMI shield over sensor and electronics, with its aperture to admit beta particles.} \label{fig:ReadoutBoard}
\end{figure}

A custom readout system with large bandwidth and minimal noise amplification was designed to read out the 3D signals. Images of the readout board can be seen in Figure \ref{fig:ReadoutBoard}. The sensors are connected to a copper pad on the readout PCB using conductive tape, and then wire bonded to a gold bond pad. The copper pad is used to bias the sensor from the backside.  It has a 1~mm diameter hole in order to admit beta particles when the system is used for coincidence measurements. The amplification system uses two stages of a GALI-S66+ monolithic Darlington pair amplifier, which have bandwidth from DC to 3 GHz with gain $\sim$ 20dB and noise figure $2.4$ dB in the frequency range of $0.1-1$~GHz \cite{GaliS66+}. The electronic components are covered by grounded electromagnetic interference (EMI) shields, with each stage of GALI-S66+ isolated in a separate shield structure. The larger EMI shield, which covers the sensor and first stage of amplification, has a 5~mm diameter aperture in order to admit the beta particles, as seen in Figure \ref{fig:ReadoutBoard2}. There is also an EMI shield covering the electronics on the backside of the board, also with an aperture. The circuit is designed with track width and clearance that match $50~\Omega$ impedance at each stage. Grounded vias surround the components and traces in order to minimize ground loops. The trace that transmits the signal to readout is a straight line, in order to minimize radio frequency emission. Four registration holes at each corner are connected to ground plates to link the board to the chassis ground. The output of the readout board is further amplified by a Particulars AM-02B (35 dB) amplifier before being read out on a Tektronix DPO7254 2.5 GHz 40 GS/s (20 GS/s in 2-channel mode) oscilloscope. A Crystek CLPFL-1000 1~GHz low-pass filter is used to reduce high-frequency noise.

Signals from the device under test (DUT) are measured in coincidence with a reference silicon sensor to suppress random coincidences. The reference sensor is an LGAD fabricated by Hamamatsu Photonics K.K.\ with excellent signal to noise ratio. The sensors, along with the mechanical framework providing precise alignment of all components, are placed and measured in an environmental chamber at $-45 \pm 1.5~^\circ$C. The low temperature was necessary to reduce noise. Taking the measurement at low temperature shifts to lower voltages both the breakdown and onset of charge multiplication by approximately the same amount. While the current in the IV data at higher temperature can be extrapolated to lower temperature, the breakdown voltage at low temperature is not reflected in the plot.

Signal waveforms from the DUT are then averaged over a large number of events and corrected for any average DC offset. The signal is integrated between the two points where the pulse crosses 0~V, to obtain the charge. More details on distinguishing the signal from noise are given below. 

\begin{figure}[htbp]
\centering
    \begin{subcaptionblock}{0.57\textwidth}
        \includegraphics[width=\textwidth]{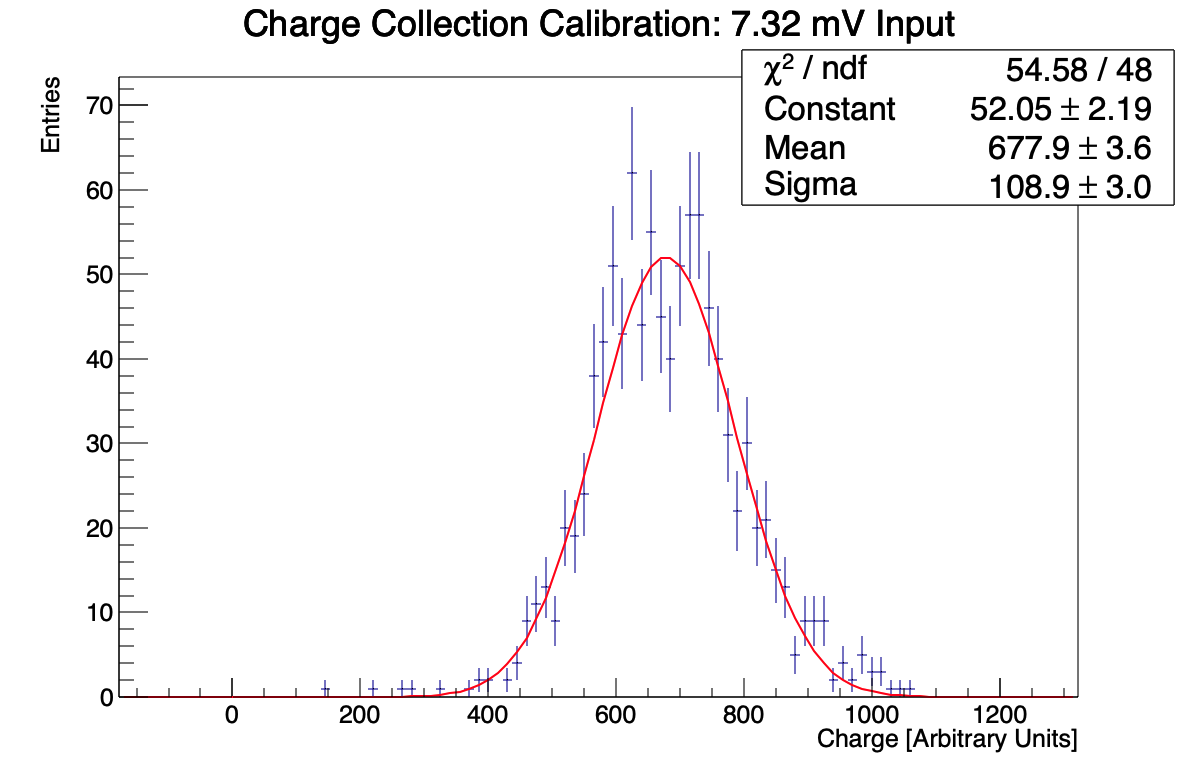}
        \caption{\label{fig:CalSingle}}%
    \end{subcaptionblock}
%\qquad
    \begin{subcaptionblock}{0.42\textwidth}
        \includegraphics[width=\textwidth]{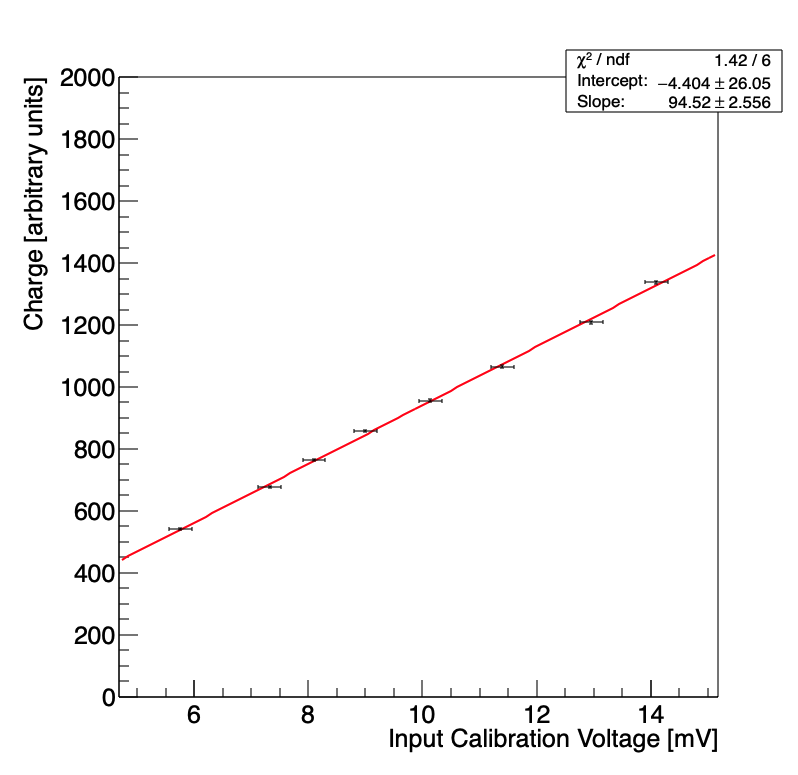}
        \caption{\label{fig:CalSlope}}%
    \end{subcaptionblock}
\caption{(a) An example Gaussian fit of the calibration data, with input voltage 7.32 mV. (b) The linear fit of arbitrary units of charge to actual charge. The slope gives the conversion factor. \label{fig:Calibration}}
\end{figure}

A calibration capacitor integrated with the readout board is used to inject a known charge for calibration. Its capacitance is measured using the CV measurement procedure and is different for each board. A typical value of its statistical measurement uncertainty is $2\%$. A Tektronix CFG250 function generator is used to pulse the capacitor with voltages in the range from about 2 to 25 mV, depending on the capacitance, and waveforms are collected and analyzed with a procedure identical to the one explained above. Using 1000 waveforms, a histogram of integrated charges is fitted with a Gaussian function, an example of which is shown in Figure \ref{fig:CalSingle}. A line can then be fitted to a graph of input charge versus the Gaussian mean, with error bars given by the fit uncertainty and the uncertainty on the measurement of the input voltage; the slope gives the conversion factor, as seen in Figure \ref{fig:CalSlope}.

This procedure was performed using five different capacitors, with values ranging from about 0.2 pF to 1.0 pF, in order to account for potential parasitic capacitance from the other components of the board, or the board itself. A line was fitted in a plot of conversion factor versus measured capacitance, with the y-intercept giving the parasitic capacitance. There is an estimated overall 3.5\% uncertainty in the calibration.

\section{Measurement Results}
\label{sec:results}

\subsection{Leakage Current and Capacitance}
\label{sec:IVCV}

\begin{figure}[h]
\centering
    \begin{subcaptionblock}{0.495\textwidth}
            \includegraphics[width=\textwidth]{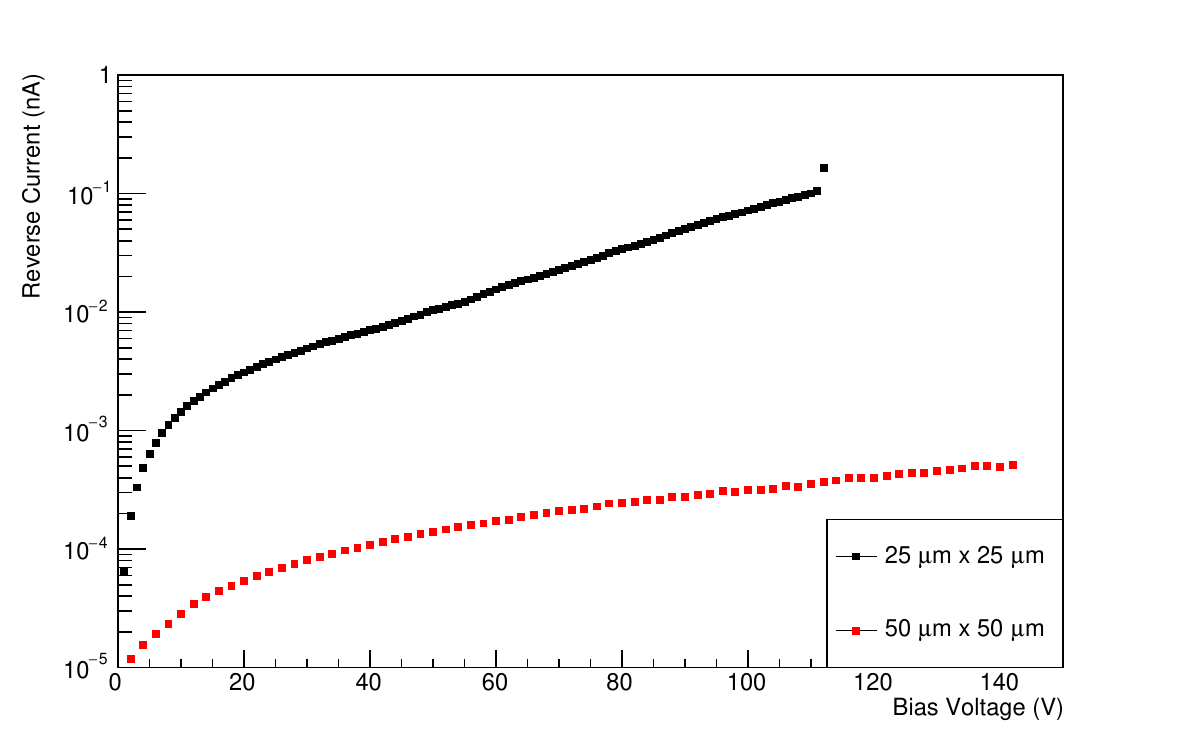}
        \caption{\label{fig:IV}}%
    \end{subcaptionblock}
%\qquad
    \begin{subcaptionblock}{0.495\textwidth}
            \includegraphics[width=\textwidth]{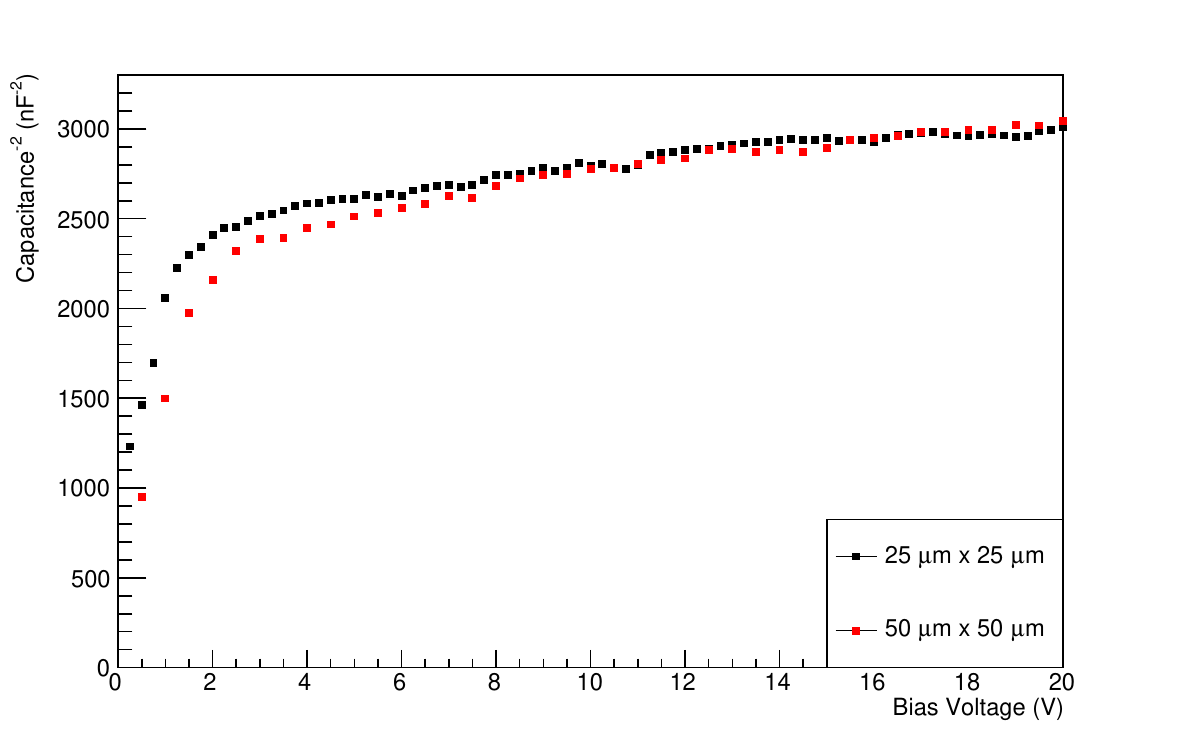}
        \caption{\label{fig:CV}}%
    \end{subcaptionblock}\caption{(a) Leakage current versus bias voltage for a sensor of pitch $25~\mu{\rm m} \times 25~\mu {\rm m}$ and one of pitch $50~\mu{\rm m} \times 50~\mu {\rm m}$, measured at 20$~^\circ$C and scaled to $-45~^\circ\mathrm{C}$. While the current can be extrapolated to $-45~^\circ$C, the breakdown voltage is not. (b) Capacitance versus bias voltage for the same sensors, measured at 20$~^\circ$C.\label{fig:IVCV}}
\end{figure}

 Figure \ref{fig:IV} shows the range of leakage current values observed in a typical sensor prior to irradiation, measured at $20~^\circ\mathrm{C}$ and scaled to $-45~^\circ\mathrm{C}$ using the conversion equation~\cite{Chilingarov:2013}

 \begin{equation}
     I(T_2) = I(T_1) \times \left(\frac{T_2}{T_1}\right)^2\exp\left(\frac{E_{\mathrm{eff}}}{2k_B}\left(\frac{1}{T_2}-\frac{1}{T_1}\right)\right).
 \end{equation}

$E_{\mathrm{eff}}$ is the silicon effective bandgap ($\sim 1.21$~eV), and $k_B$ is the Boltzmann constant. The leakage current varies between sensors at low voltage, and is typically in the range from 0.01 to 1 nA at 80 V bias. The sensor breakdown voltage ranges from about 60 V to 120 V. The latter value is in good agreement with simulations \cite{Povoli:2019} and is representative of the intrinsic breakdown in this very small pitch geometry. Small variations around this value, up to a few volts, could be due to small variations in the gap between the tips of the n-type columns and the p-type support wafer. The device is designed so that breakdown between n-type and p-type columns will occur before breakdown at the tip of the n-type column, but if the gap between the n-type column and the support wafer is slightly smaller due to intrinsic variations in the device processing, the breakdown at the tip can occur first. On the other hand, significantly smaller breakdown voltage values are likely to be ascribed to process defects, for instance, defects in the column etching that can lead to an abrupt increase of the leakage current. The inverse square of the capacitance versus bias voltage for the same sensor is also shown in Figure~\ref{fig:CV}, where we see the curve flatten around 2-3 V; this feature is observed consistently across sensors. However (see below), because the effective capacitor geometry of these devices is non-planar, at these low voltages there are zones that are not fully depleted; the depletion of these zones produces the positive slope in the plot above 3 V. A typical capacitance for $25~\mu{\rm m} \times 25~\mu {\rm m}$ and $50~\mu{\rm m} \times 50~\mu {\rm m}$ sensors is 22 pF at 10 V. While the contribution to the capacitance from the 3D columns in the $25~\mu \mathrm{m} \times 25~\mu \mathrm{m}$ sensors should be larger than in the $50~\mu \mathrm{m} \times 50~\mu \mathrm{m}$ sensors due to the smaller inter-electrode distance, the metal grid that shorts the pixels together is roughly twice as long in the $50~\mu \mathrm{m} \times 50~\mu \mathrm{m}$ ones, which causes larger capacitance for those sensors, resulting in roughly equal capacitance in either device. Both Figure~\ref{fig:IV} and Figure~\ref{fig:CV} contain corresponding plots for a $50~\mu{\rm m} \times 50~\mu {\rm m}$ sensor for comparison.

\subsection{Charge Collection}
\label{sec:CC}

Ten thousand waveforms were collected at each bias voltage. Because the charge distribution includes contributions from both genuine signals and large noise pulses, and the distributions overlap significantly, it is necessary to fit both simultaneously to extract the most probable value (MPV) of the Landau-distributed signal charges. To this end, data were collected both with and without the source in identical modes, except that there is no double coincidence required when there is no source. The data without the source (``noise data'') are fitted with a Gaussian centered at zero times a sigmoid function, which models the cutoff due to the trigger threshold. The model is {\sl ad hoc} and only meant to give a characterization of the shape of the noise distribution in order to account for instances where noise rather than a genuine MIP pulse caused the oscilloscope to trigger. The fit parameters from the noise distribution are used as a starting point for the same parameters in the fit to the shape of the data produced with the source, which is expected to be the noise sigmoid-Gaussian plus the signal Landau convolved with a Gaussian distribution. A Gaussian constraint is placed on the parameter values for the previously fitted noise distribution. The relative normalization of the two functions is a free parameter in the fit. 

The data cut off at the high end of the distribution due to limitations of the oscilloscope, for which increasing the maximum voltage would worsen the resolution in the voltage that is recorded. This condition leads to poor constraints on the convolution Gaussian standard deviation ($\sigma$ parameter). We therefore set this parameter as a constant and vary it in increments of 50~$e^-$, taking the best $\chi^2/$dof value. The effect of this on the uncertainty in the MPV is accounted for in the error analysis, discussed below. This method results in fits with $\chi^2/$dof in the range  0.8-1.6.

\begin{figure}[htbp]
\centering

    \begin{subcaptionblock}{0.495\textwidth}
        \includegraphics[width=\textwidth,trim={0 10cm 0 0},clip]{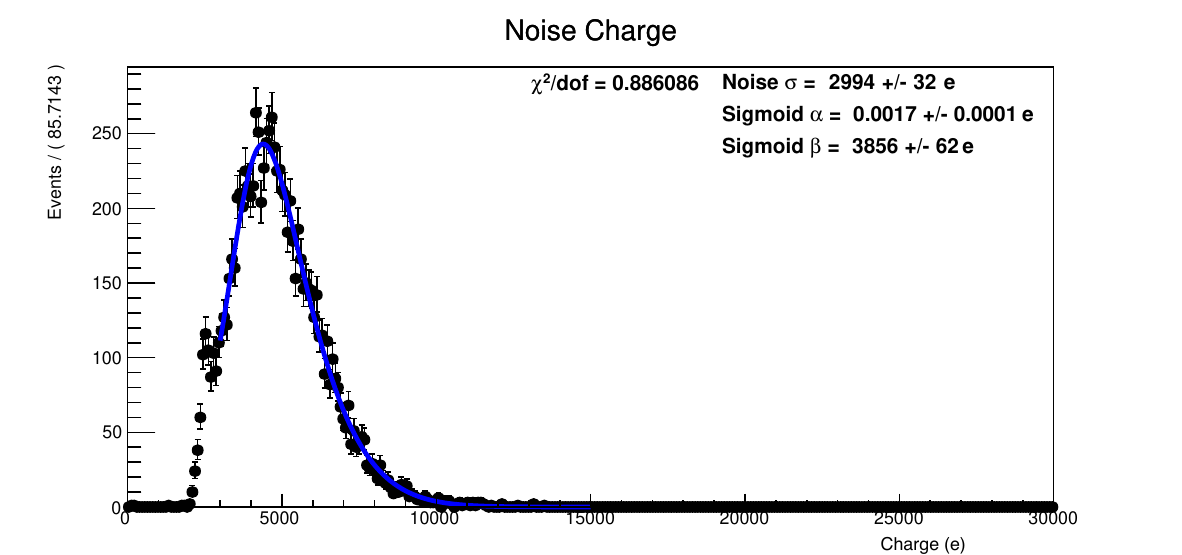}
        \caption{\label{fig:ChargeColl20Noise}}%
    \end{subcaptionblock}
    \begin{subcaptionblock}{0.495\textwidth}
        \includegraphics[width=\textwidth,trim={0 10cm 0 0},clip]{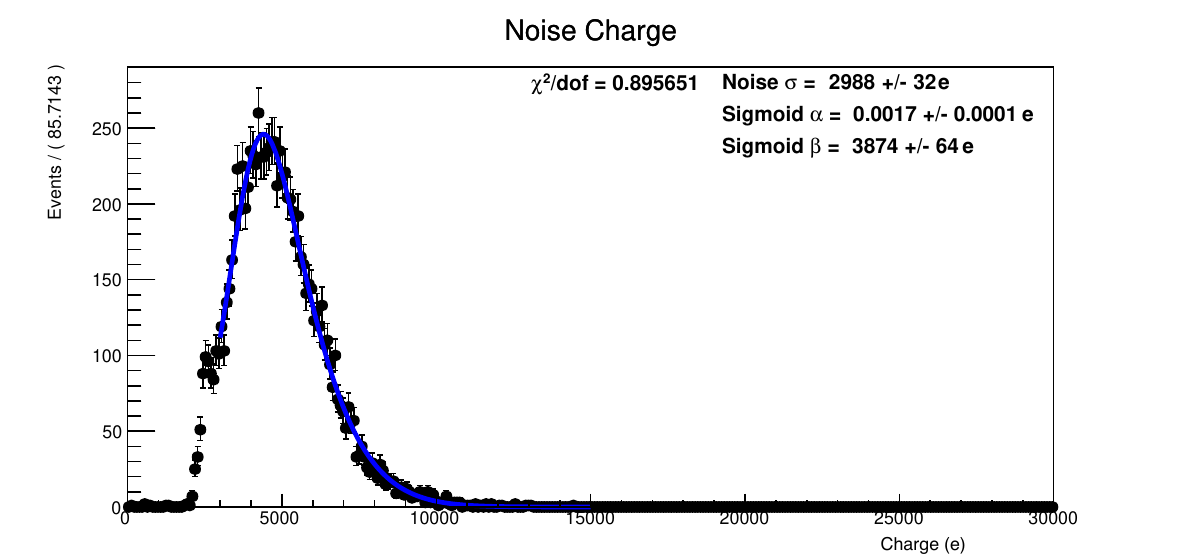}
        \caption{\label{fig:ChargeColl80Noise}}
    \end{subcaptionblock}%
    \qquad
    \begin{subcaptionblock}{0.495\textwidth}
        \includegraphics[width=\textwidth,trim={0 0 0 10cm},clip]{20VCharge2.pdf}
        \caption{\label{fig:ChargeColl20}}%
    \end{subcaptionblock}
    \begin{subcaptionblock}{0.495\textwidth}
        \includegraphics[width=\textwidth,trim={0 0 0 10cm},clip]{80VCharge2.pdf}
        \caption{\label{fig:ChargeColl80}}
    \end{subcaptionblock}%
\caption{(a) Histogram of noise distribution with fit at -20 V. (b) Histogram of noise distribution with fit at -80 V. There is little effect of leakage current on noise. (c) Histogram of charge collection for one example sensor at -20 V. (d) The same sensor's charge collection at -80 V. All are measured at -45$~^\circ$C. The fit parameters and errors are shown on the upper right corner of the plots, and the $\chi^2/$dof is shown in the upper middle portion. The three fit parameter values for the noise distributions in the upper two graphs are used as Gaussian constraints to fit the noise portion of the distributions in the respective lower two graphs, shown in green. The signal Landau is shown in red, and the sum of the two distributions fitted to the data in blue.\label{fig:ChargeColl}}
\end{figure}

Figure~\ref{fig:ChargeColl} shows graphs of noise (\ref{fig:ChargeColl20Noise}, \ref{fig:ChargeColl80Noise}), and signal plus noise (\ref{fig:ChargeColl20}, \ref{fig:ChargeColl80}), for two values of bias voltage (-20~V and -80~V) applied to one of the sensors. In Figures \ref{fig:ChargeColl20Noise} and \ref{fig:ChargeColl80Noise}, the three fit parameters for the sigmoid-Gaussian and the $\chi^2$/dof are shown in the top right. In Figures \ref{fig:ChargeColl20} and \ref{fig:ChargeColl80}, the 7 fit parameters --- 2 Landau parameters, the convolved Gaussian $\sigma$, 3 sigmoid-Gaussian parameters, and the relative normalization $f$ --- are shown in the top right of the plots, along with the $\chi^2$/dof. The noise distribution that arises in the combined fit of signal plus noise is shown in green, and the Landau convolved with a Gaussian is shown in red, with the sum in blue. 

It is crucial to compute the integral of each waveform rather than assume that the peak height is proportional to the charge. The noise pulses tend to be faster and therefore have a smaller integral than real signal pulses. In a histogram of peak heights, the noise and signal are indistinguishable, and only when the integration is performed is a significant separation seen between signal and noise peaks, as can be seen in Figures \ref{fig:ChargeColl20} and \ref{fig:ChargeColl80}.

\begin{figure}[h]
\centering
    \begin{subcaptionblock}{0.495\textwidth}
        \includegraphics[width=\textwidth,trim={0 0 0 1cm},clip]{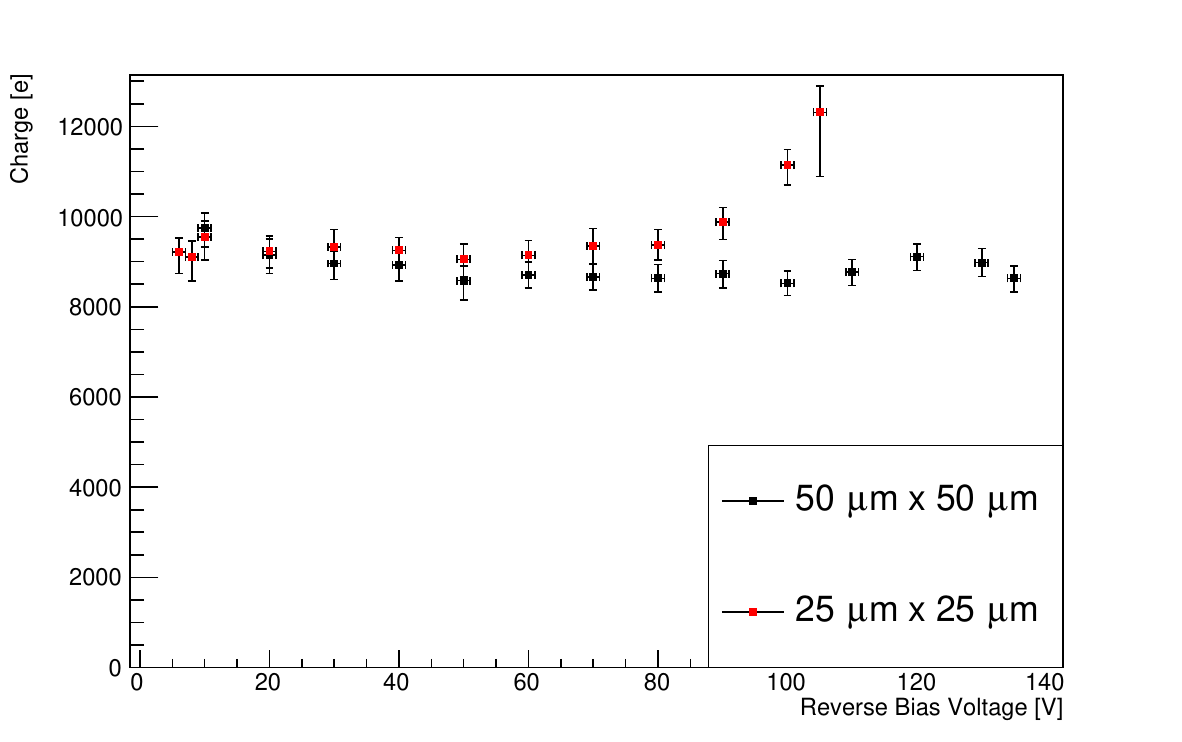}
        \caption{\label{fig:ChargeCollSum}}%
    \end{subcaptionblock}
    \begin{subcaptionblock}{0.495\textwidth}
        \includegraphics[width=\textwidth,trim={0 0 0 1cm},clip]{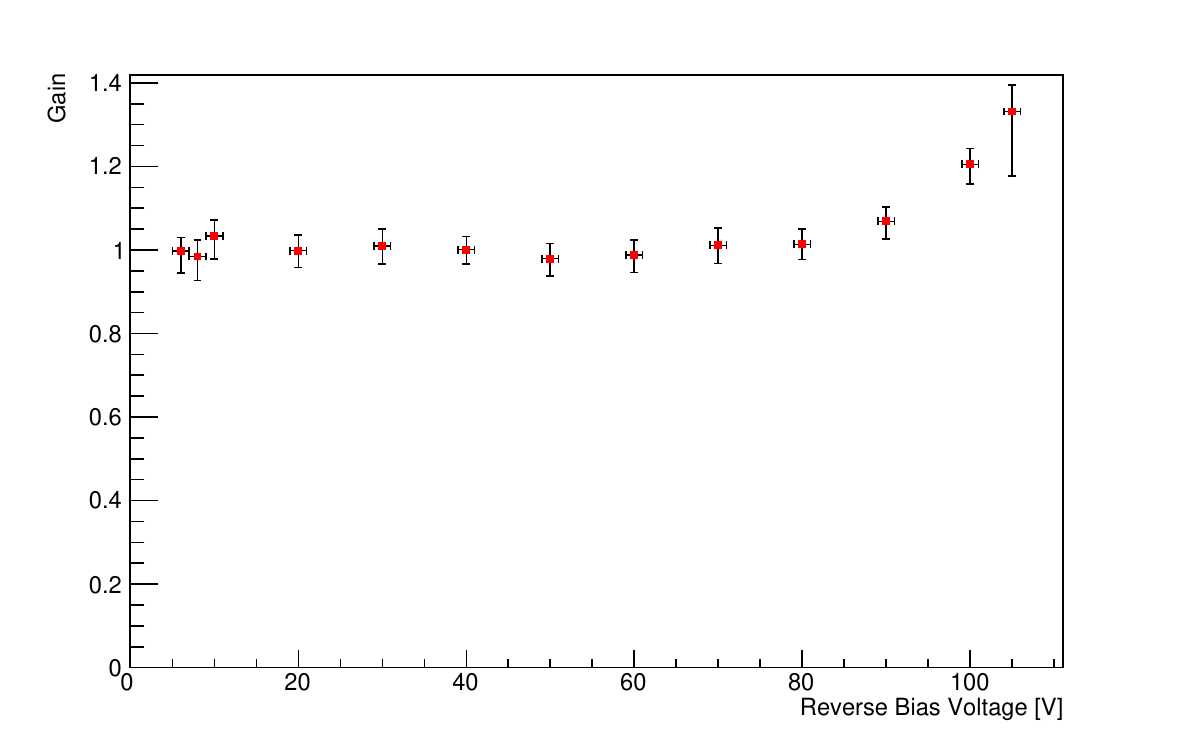}
        \caption{\label{fig:ChargeCollGain}}%
    \end{subcaptionblock}
\caption{(a) The most probable value of the charge collected versus bias voltage for example $25~\mu{\rm m}\times 25~\mu{\rm m}$ pitch and $50~\mu{\rm m}\times 50~\mu{\rm m}$ pitch 3D sensors, measured at -45$~^\circ$C. (b) The gain of the $25~\mu{\rm m}\times 25~\mu{\rm m}$ sensor calculated by dividing by the average between bias values from -20 to -80~V. The data suggest there is gain in $25~\mu {\rm m}\times 25~\mu {\rm m}$ sensors starting at about 90 V. The error bars represent the statistical and systematic uncertainties added in quadrature. There is an additional 3.5\% error on the overall calibration of the charge, which is not included in the error bars.  \label{fig:ChargeCollSumFull}}
\end{figure}

The MPV versus bias voltage for this sensor for bias voltages from 6~V to 105~V (just below breakdown) is shown in Figure \ref{fig:ChargeCollSum}. For comparison, measurements were taken of identically designed sensors with $50~\mu {\rm m}\times 50~\mu$m pitch. Data for one $50~\mu {\rm m}\times 50~\mu$m sensor can also be seen in Figure \ref{fig:ChargeCollSum}.

The nominal thickness of the sensors is $150~\mu$m; however the actual active thickness may be less due to boron diffusion from the support wafer. A collected charge of about 9400 $e^-$ is consistent with an active thickness of about 140 $\mu$m, assuming 67 electron-hole pairs/$\mu$m for MIP ionization \cite{ParticleDataGroup:2022pth,Weigell:2012gfa}.

We see a gradual decrease in the relative normalization (parameter $f$ in the statistics of Figure~\ref{fig:ChargeColl}) as the bias voltage increases, from $0.27 \pm 0.01$ at the lowest voltage to $0.13 \pm 0.01$ at 100 V. The proposed explanation derives from the low-field zones in 3D sensors \cite{Parker:1996dx}, and that as bias increases, more of the central region of each $25~\mathrm{\mu m} \times 25~\mathrm{\mu m}$ cell is depleted; at low voltages the pairs created by MIPs passing through low field zones drift slowly; these regions are subsequently depleted at higher voltages. It will be interesting to study this further with TCT and beam tests of sensor efficiency versus bias voltage. By 105~V, a larger proportion of signals are missed because they are too large for the oscilloscope's range. 

The data are not significantly noisier at higher voltages; the noise distributions are nearly identical from 6 V up to 100 V. While it is expected that irradiated sensors will have increased leakage current and somewhat higher noise, with these sensors the current is small enough at $-45~^\circ$~C to be a sub-dominant noise source even up to the highest voltage below breakdown.

Amplification of the signal in a $25~\mu {\rm m}\times 25~\mu$m sensor is seen in Figure \ref{fig:ChargeCollGain} starting at 90 V bias, and the signal is multiplied by a factor of 1.33 at 105 V. The gain is calculated by dividing the MPV at each voltage by the average of the MPV's between -20 and -80 V bias. Amplification is not predicted nor observed in $50~\mu {\rm m}\times 50~\mu{\rm m}$ sensors up to their breakdown voltage. The gain can be attributed to impact ionization charge multiplication, due to the high electric fields in the compact geometry of the detector. Charge multiplication has been reported in strip sensors of a larger pitch~\cite{Kohler:2011zzc}; the gain observed here is larger and onsets at a lower voltage. The trend of multiplication starting at about 90 V is consistently observed across the set of $25~\mu {\rm m}\times 25~\mu$m test structures.

\subsection{Error Analysis}
\label{sec:Error}

Each IV and CV measurement point shown in Figures \ref{fig:IV} and \ref{fig:CV} is the average of three measurements, and the standard deviation in all cases is less than 2\%. The primary contributions to systematic uncertainties are those associated with the setup configuration (1.9\%), the accuracy and precision of the source and measurement equipment ($\pm 0.3\% + 100$ fA for the Keithley 237; $\pm 0.029\% + 300$~pA for the Keithley 2410, and $\pm 0.34 \%$ for the HP4284A), and the precision of the measurement of the temperature ($\pm 0.5~^\circ$C leads to an uncertainty of $\pm 1.82\%$ on leakage current). 

Three kinds of error were assessed for the charge collection measurement: statistical error in the fit, systematic error due to constraints on the standard deviation of the Gaussian function, and systematic error due to the choice of oscilloscope trigger threshold.

The statistical error is assessed by dividing the 10,000 waveforms into subsets and fitting each subset in the same manner as was previously described. The pure noise histogram is not refitted each time. The standard deviation of all of the most probable values divided by the square root of the number of subsets is taken to be the statistical error. This value is seen to remain consistent regardless of the number of subsets.

As mentioned above, the convolved Gaussian $\sigma$ parameter is set at a constant value. It is observed that in the range of $\sigma$ values from 100 to 1000 $e^-$, the $\chi^2/$dof changes by $\lesssim$ 10\%.  This variation is due to the upper tail of the Landau that is convolved with a Gaussian being cut off and to the lower tail being overlapped with the noise. A small $\sigma$ leads to a larger MPV and vice versa. The value of the constant $\sigma$ is varied in increments of $50~e^-$ and the best fit value is used.  The range for which the $\chi^2/$dof is within 10\% of the best value is taken as the error range associated with the Gaussian $\sigma$ systematic effects.

To assess the error due to the choice of trigger threshold, the trigger threshold was varied and data were collected across the range of thresholds for one applied voltage. After accounting for variation due to statistical error, an additional 3\% uncertainty is associated with the trigger threshold.

The uncertainty associated with the calibration is related to knowledge of the calibration capacitance. The measurement of the calibration capacitance is repeated several times, and the standard deviation gives an error of 2\%. For the information reported in Figure \ref{fig:ChargeCollSum}, the calibration capacitance is measured to be $0.299 \pm 0.005$ pF. Additionally the parasitic capacitance is $0.016 \pm 0.010$~pF. That uncertainty comes from the linear fit of capacitance versus calibration slope. Taken together the calibration capacitance is $0.315 \pm 0.011$ pF, which is 3.5\% uncertainty. This added uncertainty is not reflected on the error bars in Figure \ref{fig:ChargeCollSum}, since it is the same for each data point.

\section{Conclusions}
\label{sec:Conc}

Leakage current and breakdown voltage within acceptable operational ranges are observed for a set of unirradiated 3D silicon pixel detectors with one readout electrode per pixel (1E) and pitch $25~\mu {\rm m} \times 25~\mu {\rm m}$. Measurement of bulk capacitance versus applied bias indicate substantial depletion of the bulk below 1~V, and then further gradual depletion of the low field regions as bias increases up to about 10~V. Charge collection efficiency is observed to be {\textbf consistent with that of a $50~\mu{\rm m} \times 50~\mu{\rm m}$} sensor. Gain up to 1.33 is seen beginning at about 90~V bias for an unirradiated sensor, which is below the breakdown voltage and consistent with simulations; this is promising for the potential of operating a thin small-pitch 3D sensor with geometrical internal gain in order to compensate for lost charge collection efficiency. Post-irradiation data for these devices will be discussed in a subsequent companion publication.

\acknowledgments

This work was supported by U.S. Department of Energy grant DE-SC0020255 and Federal Award Number 80NSSC20M0034 (2020-RIG) from the National Aeronautics and Space Administration (NASA). This work was further supported by the Italian National Institute for Nuclear Physics (INFN), 1st Scientific Commission (CSN1) with
Projects RD\_FASE2 and FASE2\_ATLAS; it was also supported by the AIDA-2020 project EU-INFRA proposal no.~654168. The authors are grateful for the support of Prof.~Paulo Oemig of New Mexico State University and of Dr.~Regina Caputo of NASA Goddard Space Flight Center.

% Bibliography

%% [A] Recommended: using JHEP.bst file
\bibliographystyle{JHEP}
\bibliography{main.bib}

\appendix

%% or
%% [B] Manual formatting (see below)
%% (i) We suggest to always provide author, title and journal data or doi:
%% in short all the informations that clearly identify a document.
%% (ii) please avoid comments such as "For a review'', "For some examples",
%% "and references therein" or move them in the text. In general, please leave only references in the bibliography and move all
%% accessory text in footnotes.
%% (iii) Also, please have only one work for each \bibitem.

% \begin{thebibliography}{99}

% \bibitem{a}
% Author,
% \emph{Title},
% \emph{J. Abbrev.} {\bf vol} (year) pg.

% \bibitem{b}
% Author,
% \emph{Title},
% arxiv:1234.5678.

% \bibitem{c}
% Author,
% \emph{Title},
% Publisher (year).

% \end{thebibliography}
\end{document}